\documentclass[a4paper]{article}

\usepackage{INTERSPEECH2021}
\usepackage{hyperref}
\usepackage{multirow}
\usepackage{flushend}

\title{Integrating Frequency Translational Invariance in TDNNs and Frequency Positional Information in 2D ResNets to Enhance Speaker Verification}
\name{Jenthe Thienpondt, Brecht Desplanques, Kris Demuynck}

\address{
    IDLab, Department of Electronics and Information Systems, Ghent University - imec, Belgium
}
\email{jenthe.thienpondt@ugent.be, brecht.desplanques@ugent.be}

\begin{document}

\maketitle
\begin{abstract}
This paper describes the IDLab submission for the text-independent task of the Short-duration Speaker Verification Challenge 2021 (SdSVC-21). This speaker verification competition focuses on short duration test recordings and cross-lingual trials, along with the constraint of limited availability of in-domain DeepMine Farsi training data. Currently, both Time Delay Neural Networks (TDNNs) and ResNets achieve state-of-the-art results in speaker verification. These architectures are structurally very different and the construction of hybrid networks looks a promising way forward. We introduce a 2D convolutional stem in a strong ECAPA-TDNN baseline to transfer some of the strong characteristics of a ResNet based model to this hybrid CNN-TDNN architecture. Similarly, we incorporate absolute frequency positional encodings in an SE-ResNet34 architecture. These learnable feature map biases along the frequency axis offer this architecture a straightforward way to exploit frequency positional information. We also propose a frequency-wise variant of Squeeze-Excitation (SE) which better preserves frequency-specific information when rescaling the feature maps. Both modified architectures significantly outperform their corresponding baseline on the SdSVC-21 evaluation data and the original VoxCeleb1 test set. A four system fusion containing the two improved architectures achieved a third place in the final SdSVC-21 Task 2 ranking.

\end{abstract}

\section{Introduction}

Speaker verification determines whether the speaker in a test utterance matches with a claimed enrolled speaker. Current state-of-the-art speaker verification systems use a deep neural network trained on utterance classification according to speaker identity. The current most popular neural network topologies are based on Time Delay Neural Network (TDNN) \cite{x_vectors, ecapa_tdnn}, also referred to as x-vectors, and ResNet \cite{magneto} architectures. The converged network is subsequently used to extract low-dimensional speaker embeddings from a bottleneck layer in the final part of the network. The speaker similarity is expressed by the cosine distance between the speaker embeddings.


In this paper we propose architectural extensions to the current state-of-the-art ECAPA-TDNN and ResNet based speaker embedding extractors. We enhance the ECAPA-TDNN~\cite{ecapa_tdnn} architecture with a network stem based on 2D convolutions. This allows the network to initially construct local, frequency-invariant features before applying 1D convolutions which explicitly integrate the frequency positional information of the features. We also propose the addition of absolute positional frequency encodings in the ResNet based architectures. These learnable feature map offsets enable the ResNet architecture to exploit frequency positional information. Additionally, we introduce a frequency-wise Squeeze-Excitation (SE) block. This module improves upon the standard SE-block \cite{se_block} by enforcing the global utterance information that is injected in the local convolutions to be estimated per frequency component.

The proposed modifications are evaluated in the context of the text-independent speaker verification Task 2 of the Short-duration Speaker Verification Challenge 2021 (SdSVC-21)~\cite{sdsvc_21}. Building upon our previously introduced large margin fine-tuning procedure \cite{icassp_voxsrc20}, we evaluate a similar fine-tuning strategy to adapt the models to the in-domain Farsi speech of this challenge. Finally, we use a quality-aware score calibration stage~\cite{icassp_voxsrc20} that uses duration- and language-based quality measurements to generate speaker similarity scores that are more consistent across the varying trial conditions.




\section{Baseline system architectures}

\subsection{ECAPA-TDNN}

The ECAPA-TDNN \cite{ecapa_tdnn} architecture is an enhanced version of the popular x-vector topology \cite{x_vectors, x_vector_wide}.
The use of hierarchical grouped convolutions~\cite{res2net} reduces the model parameter count. It also introduces 1-dimensional TDNN-specific SE-blocks~\cite{se_block} which rescale the intermediate time context bound frame-level features per channel according to global utterance properties.
The pooling layer uses a channel-and context-dependent self-attention mechanism to attend to different speaker characterizing properties at different time steps for each feature map.
Finally, Multi-layer Feature Aggregation (MFA)~\cite{multi_stage_aggregation} provides additional complementary information for the statistics pooling by concatenating the final frame-level features with the intermediate features of previous layers. The architecture is trained with the AAM-softmax \cite{arcface} loss. More details about the architecture can be found in~\cite{ecapa_tdnn}. Similar to~\cite{icassp_voxsrc20}, we increase the intermediate channel dimension to 2048 and add a fourth SE-Res2Block to the network.

\subsection{SE-ResNet34}

Our second baseline system is based on the ResNet \cite{resnet} architecture. We use the same network topology as defined in~\cite{magneto}, with the addition of an SE-block at the end of each residual block. We also incorporate the same attentive statistics pooling layer as the ECAPA-TDNN baseline. A final modification is the incorporation of sub-center AAM~\cite{subcenter}. This is an extension of the AAM-softmax loss function and it defines multiple prototype embeddings per training speaker. This should make the model less susceptible to potential label noise or severely corrupted data in the training set. In this paper the number of sub-centers per speaker is set to two. This system was the best scoring single system on the VoxSRC-20 validation set in our submission~\cite{voxsrc_2020_technical_report}.

\subsection{Standard speaker embedding training procedure}

All baseline neural networks are trained on all of the allowed training corpora in the SdSVC-21 setting. The input features are 80-dimensional log Mel-filterbank energies extracted with a window length of 25 ms and a frame-shift of 10 ms. We apply an online augmentation strategy on the 2\,s random training crops using the MUSAN \cite{musan} corpus (additive music, noise \& babble) and the RIR \cite{rirs} corpus (reverb). SpecAugment \cite{specaugment} is applied to the input features with a frequency and temporal masking dimension of 10 and 5, respectively. Finally, the input features are mean normalized across time per utterance.

The model parameter updates are determined by the Adam optimizer \cite{adam} with a cyclical learning rate schedule. We use the \textit{triangular2} policy described in \cite{clr}. The cycle length is set to 130k iterations with a minimum and maximum learning rate of 1e-8 and 1e-3, respectively. The systems are trained for three full cycles. To prevent overfitting, a weight decay of 2e-4 is applied on the weights of the AAM-softmax layer, while a value of 2e-5 is used on all other layers of the model.

\section{Proposed system architectures}

TDNN and ResNet based speaker verification systems achieve similar performance~\cite{voxsrc_2020_technical_report} while being structurally very different. TDNNs depend on 1D convolutions of which the kernels cover the complete frequency range of the input features. As a consequence, the absolute frequency position of each input feature is hard-coded through the order of the filter coefficients. This makes sense in the context of speaker verification as features based on absolute frequency information such as the speaker's pitch provide crucial information. The main drawback is that many filters are needed to model the fine details of complex patterns that can occur at any frequency region. 


ResNets use 2D convolutions with a small receptive field, capturing local frequency and temporal information. By exploiting local speaker-specific frequency patterns that repeat but for small frequency shifts, this model requires fewer feature channels to model fine frequency details. However, accurate absolute frequency positions of features are not explicitly encoded~\cite{position_info_resnet}. This can be sub-optimal as we expect significantly different patterns in the low frequency regions compared to the high frequency regions. 
In the next sections we enhance both architectures to incorporate the beneficial characteristics of the other model.

\begin{figure}[ht]
\centering
\includegraphics[scale=0.32]{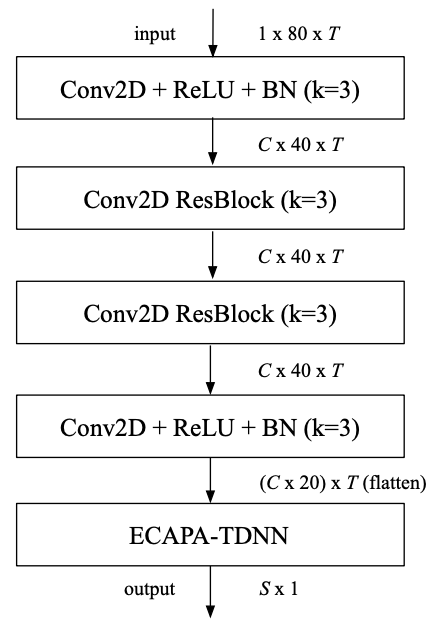}
\caption{{\it The 2D convolutional stem of the ECAPA CNN-TDNN architecture. $C$ and $T$ indicate the channel and temporal dimension, respectively. The kernel size is denoted by $k$. $S$ refers to the number of training speakers.}}
\label{fig:conv_stem}
\end{figure}

\subsection{ECAPA CNN-TDNN with 2D convolutional stem}

We want the ECAPA-TDNN to be invariant to small and reasonable shifts in the frequency domain, compensating for realistic intra-speaker frequency variability. To accomplish this, we base the initial network layers on 2D convolutions. These layers will also require less filters to model high resolution input detail. A similar modification of the standard x-vector network for text-dependent speaker verification can be found in~\cite{NetEase_CNN_TDNN}.

Inspired by the powerful 2D convolutions in the ResNet architectures, we adopt a similar structure for our ECAPA-TDNN stem. The proposed network configuration is shown in Figure~\ref{fig:conv_stem}. The number of channels $C$ in the stem is set to 128 for all experiments. To make the stem more computationally efficient we use a stride of 2 in the frequency dimension of the first and final 2D convolutional layer. The output feature map of the new stem is subsequently flattened in the channel and frequency dimensions and used as input for the regular ECAPA-TDNN network. We will refer to this network as the ECAPA CNN-TDNN.

\subsection{SE-ResNet34 with frequency position information}

\subsubsection{Learnable frequency positional encodings}

While a certain robustness against frequency translation is beneficial, the spatial invariance of 2D convolutions can limit the ability of the network to fully exploit frequency-specific information. We argue that encoding positional frequency information in the intermediate feature representations can make the network incorporate and utilize knowledge of the feature frequency positions.

Positional encodings have been popularized with the rise of Transformer~\cite{transformer} models. By design, these Transformers are invariant to the re-ordering of the input sequence and positional encodings can alleviate this issue when required. These encodings can either be learnable or use a fixed representation based on the sine and cosine function \cite{seq_to_seq_fb, transformer}.
In this paper we focus on learnable encodings as the computational overhead is negligible.

Consider the input of a residual module in our baseline ResNet architecture as $ \pmb{X} \in \mathbb{R}^{C \times F \times T} $,  with $C$, $F$ and $T$ indicating the channel, frequency and time dimension, respectively.
We define the positional encoding vector $\pmb{p} \in \mathbb{R}^{F} $ as a trainable vector.
The elements in this vector are broadcasted to match the dimensions of the targeted input feature map. The input of the residual module is now defined as $\pmb{X} + \pmb{p}$. We add a unique positional encoding to the input of each residual block after branching the skip connection. 


\subsubsection{Frequency-wise Squeeze-Excitation (fwSE)}

Squeeze-Excitation (SE)~\cite{se_block} blocks have been successfully applied in both TDNN and ResNet based speaker verification architectures. The first stage, the squeeze operation, calculates a vector $ \pmb{z} \in \mathbb{R}^{C} $ containing the mean descriptor for each feature map. It is followed by the excitation operation, which calculates a scalar for each feature map given the information in $\pmb{z}$. Subsequently, each feature map is rescaled with its corresponding scalar.

We argue that rescaling per feature map is not tailored for the processing of speech in ResNets. Instead, we propose a frequency-wise Squeeze-Excitation (fwSE) block, which injects global frequency information across all feature maps. We calculate a vector $ \pmb{z} \in \mathbb{R}^{F} $ containing the mean descriptor for each frequency-channel of the intermediate feature maps in the following manner:
\begin{equation}
\label{channel_se}
z^{f} = \frac{1}{C \times T} \sum_{i=1}^{C} \sum_{j=1}^{T} x^{f}_{i,j}
\end{equation}
with $x^f$ the elements of $\pmb{X}^{f} \in \mathbb{R}^{C \times T}$, the component of input feature map $ \pmb{X} \in \mathbb{R}^{C \times F \times T} $ corresponding with frequency position $f$.
From the mean descriptors in $ \pmb{z} $ we calculate a vector $ \pmb{s} \in \mathbb{R}^F $ containing the scaling scalars for each frequency-channel in the second stage, the excitation operation:
\begin{equation}
\label{se_excitate}
\pmb{s} = \sigma( \pmb{W}_2 f(\pmb{W}_1 \pmb{z} + \pmb{b}_{1}) + \pmb{b}_2)
\end{equation}
with $ \pmb{W} $ and $ \pmb{b} $ indicating the weights and bias of a linear layer, $ f(.) $ the ReLU activation function and $ \sigma $ the sigmoid function. Finally, $\pmb{X}^f$ is scaled with the corresponding scalar in $\pmb{s}$. The proposed frequency-wise SE-blocks are inserted at the end of each residual module before the additive skip connection.


\begin{table*}
  \caption{EER and MinDCF of the baseline and proposed architectures on VoxCeleb1 and SdSVC-21 validation and test sets.}
  \label{tab:exp_results}
  \centering
  \begin{tabular}{ccccccccc}
    \toprule
    \multicolumn{1}{c}{\textbf{Architecture}} & 
    \multicolumn{1}{c}{\textbf{Variant}} &
    \multicolumn{3}{c}{\textbf{VoxCeleb1-O}} &
    \multicolumn{2}{c}{\textbf{SdSVC-21 Val}} &
    \multicolumn{2}{c}{\textbf{SdSVC-21 Test}} \\
    \cmidrule(lr){3-5} \cmidrule(lr){6-7} \cmidrule(lr){8-9}
    \multicolumn{2}{c}{\textbf{}} &
    \multicolumn{1}{c}{\textbf{EER(\%)}} & \multicolumn{1}{c}{\textbf{MinDCF}} & \multicolumn{1}{c}{\textbf{MinDCF2}} &
    \multicolumn{1}{c}{\textbf{EER(\%)}} & \multicolumn{1}{c}{\textbf{MinDCF}} &
    \multicolumn{1}{c}{\textbf{EER(\%)}} & \multicolumn{1}{c}{\textbf{MinDCF}} \\
    \midrule
      & Standard & 0.85 & 0.0498 & 0.1227 & 1.86 & 0.0783 & 1.35 & 0.0632 \\
     ECAPA-TDNN & LM-FT & 0.76 & 0.0481 & 0.1010 & 1.64 & 0.0631 & 1.20 & 0.0548 \\
      & LM-FT + QMs & 0.73 & 0.0466 & 0.0961 & 1.50 & 0.0625 & 1.14 & 0.0520 \\
    \midrule
      & Standard & 0.72 & 0.0477 & 0.1193 & 1.43 & 0.0683 & 1.21 & 0.0568 \\
      ECAPA CNN-TDNN & LM-FT & 0.66 & 0.0391 & \textbf{0.0820} & \textbf{1.43} & 0.0664 & 1.08 & 0.0502 \\
      & LM-FT + QMs & \textbf{0.61} & \textbf{0.0372} & 0.0903 & 1.45 & \textbf{0.0584} & \textbf{1.01} & \textbf{0.0471} \\
    \midrule
    \midrule
      & Standard & 0.78 & 0.0432 & 0.1008 & 1.70 & 0.0872 & 1.39 & 0.0601 \\
      SE-ResNet34 & LM-FT & 0.70 & 0.0428 & 0.0884 & 1.57 & 0.0720 & 1.22 & 0.0534 \\
      & LM-FT + QMs & 0.65 & 0.0389 & 0.0877 & 1.43 & 0.0653 & 1.17 & 0.0510 \\
    \midrule
      \multirow{3}{*}{\shortstack[c]{fwSE-ResNet34\\w/ positional\\ encodings}} & Standard & 0.70 & 0.0410 & 0.0856 & 1.59 & 0.0776 & 1.23 & 0.0526 \\
      & LM-FT & 0.63 & 0.0354 & \textbf{0.0776} & \textbf{1.36} & 0.0652 & 1.13 & 0.0483 \\
      & LM-FT + QMs & \textbf{0.61} & \textbf{0.0335} & 0.0812 & 1.43 & \textbf{0.0632} & \textbf{1.08} & \textbf{0.0463} \\
    \bottomrule
  \end{tabular}
\end{table*}


\section{SdSVC-21 specific modifications}

In this section we describe the modifications to our speaker verification pipeline for the Short-duration Speaker Verification Challenge 2021 (SdSVC-21). The challenge creates a difficult speaker verification set by incorporating short duration cross-lingual trial pairs. All speakers in the set are native Farsi speakers with the trial test utterances containing either Farsi or English speech. The allowed training data consists of VoxCeleb1 \& VoxCeleb2 \cite{vox1, vox2}, LibriSpeech, from which we only use the \textit{train-other-500} component \cite{libri}, the Farsi subset of the Mozilla Common Voice corpus \cite{cv_farsi} and a part of the DeepMine dataset \cite{deepmine} with utterances from 588 in-domain speakers. As opposed to last year’s challenge, the in-domain training data is extended with English spoken utterances. More information about the challenge can be found in the SdSVC-21 evaluation plan \cite{sdsvc_21}.

\subsection{Domain-balanced large margin fine-tuning}
\label{sec:lm_ft}

We fine-tune all systems with an adapted version of the Large Margin Fine-Tuning (LM-FT) strategy presented in \cite{icassp_voxsrc20}. The performance gains of regular LM-FT were much smaller on the in-domain SdSVC-21 validation set compared to the other domains in the training set. 
To mitigate this issue, we propose a domain-balanced and SdSVC-21 specific version of the LM-FT strategy.

To preserve the verification performance on the shortest SdSVC-21 test utterances, we increase the duration of the temporal crop to only 3\,s during LM-FT. To compensate, we increase the AAM-softmax margin to a less aggressive value of 0.3. In comparison, we used an increase to 6 s and a margin of 0.5 on VoxCeleb data previously~\cite{icassp_voxsrc20}. We also equalize the sampling probability of each speech domain present in our training dataset. We sample a random utterance for each of the 588 in-domain DeepMine speakers and for each speaker in a random subset of 588 speakers from each of the other domains. Once the pool of selected utterances is exhausted during batch creation, the random selection of training samples is repeated. Due to time constraints, we disabled the hard prototype mining based utterance selection criterion~\cite{sdsvc_paper, icassp_voxsrc20}.

\subsection{Score estimation and normalization}
\label{sec:score_norm}

The speaker enrollment models are constructed by averaging the corresponding $L_2$-normalized enrollment embeddings. The verification trials are scored by calculating the cosine distance between the enrollment model and the test utterance embedding. Scores are normalized using top-2000 adaptive s-normalization \cite{as_norm_original_1, as_norm_original_2}. The imposter cohort consists of top imposters selected from the pooled SdSVC-21 Farsi training data and the Farsi component of Common Voice. Each imposter speaker is represented by the average of their length-normalized training embeddings. 

\subsection{Score fusion and quality-aware score calibration}

Score calibration maps the trial scores to log-likelihood-ratios that can be converted to interpretable probabilities~\cite{bosaris}. It also allows for easy score fusion by producing a weighted average across system scores~\cite{bosaris}. In addition, the score calibration stage can be used to compensate for variability in trial conditions by including Quality Measurements (QMs) \cite{bosaris, icassp_voxsrc20} such as duration metrics. This is very effective for evaluation sets with varying duration conditions, since most of the embedding extractors are trained with fixed-length audio crops.


The calibration is based on logistic regression with learnable weights and bias to scale and shift the original output score together with the QMs to obtain a calibrated log-likelihood-ratio~\cite{icassp_voxsrc20}. This procedure makes the evaluation decision thresholds implicitly condition dependent. To train the parameters we create our own speaker verification trials from the in-domain training set. We generate 100 cross-gender verification trials per training speaker. We select between 1 and 10 Farsi enrollment utterances per trial. The number of target and non-target trials as well as the number of Farsi-only and cross-lingual trials is balanced.





For our fusion submission, we first take a weighted average according to the performance of the systems on the SdSVC-21 validation set. Subsequently, we apply the quality-aware calibration on these averaged scores. The logistic regression uses the quality metrics described below.


\subsubsection{Duration-based quality measures}

The SdSVC-21 trials are asymmetric in the sense that there is significantly more enrollment speech than test speech. 
We introduce $\log(d_t - d_{min})$ as a QM with $d_t$ the duration of the test utterance and $d_{min}$ the minimum expected test utterance duration in the DeepMine corpus (i.e.\ 1 s).

It is harder to anticipate the impact of the multiple enrollment durations on the embedding quality.
We introduce $\log(n_e)$ as a basic second QM, where $n_e$ is the number of enrollment utterances associated with the trial. We cap $n_e$ to a maximum of three, as we expect the quality improvement on the average enrollment embedding to saturate.

\subsubsection{Cross-lingual likelihood-based quality measures}

Due to the availability of in-domain English training speech in SdSVC-21, the performance impact of cross-lingual trials seems rather limited. We notice small gains by introducing the language log-likelihood-ratio QM on the test utterance produced by a Farsi vs.\ English language classifier. The language identification system is a Gaussian Backend (GB) trained on the output of the neural network pooling layer on the SdSVC-21 training set. The GB achieves 99.8\% language classification accuracy on the SdSVC-21 validation test utterances.

\section{Results}

 We evaluate the systems on the original VoxCeleb1 test set and the SdSVC-21 validation and test set. We report the EER and MinDCF metric using a $P_{target}$ value of $10^{-2}$ with $C_{Miss} = 10$ and $C_{FA} = 1$, as this is the main metric for SdSVC-21. For VoxCeleb1-O, we also report the MinDCF2 value with $C_{Miss} = 1$. All reported scores are s-normalized according to Section~\ref{sec:score_norm} for the SdSVC-21 results. For scores on the VoxCeleb1-O trials, we use the training part of the VoxCeleb1 dataset to calculate the imposters in the s-norm cohort with a size of~120. For system discussions we only focus on the large SdSVC-21 test set due to its size and thus reliability.
 
Table~\ref{tab:exp_results} compares the performance of the baseline models with the proposed architectures. It also shows the impact of our proposed domain-balanced LM-FT strategy and of the calibration with the proposed QMs. We use a margin value of 0.3 and a crop size of 3\,s for all systems during the LM-FT stage as this provides the best performance on the in-domain SdSVC-21 sets as shown in Table~\ref{tab:ft_results}. This table also includes a fine-tuning strategy that only modifies the sampling to be domain-balanced. For VoxCeleb1-O trials we only include the duration-based QMs as defined in \cite{voxsrc_2020_paper}. 

\begin{table}[h]
    \caption{Evaluation of the domain-balanced large margin fine-tuning strategy on the ECAPA CNN-TDNN architecture.}
    \label{tab:ft_results}
  \centering
  \begin{tabular}{lccc}
    \toprule
        \multicolumn{1}{c}{\textbf{Method}} &
        \multicolumn{1}{c}{\textbf{Vox1-O}} &
        \multicolumn{2}{c}{\textbf{SdSVC-21 Test}} \\
        \cmidrule(l){2-2} \cmidrule(lr){3-4}
    \textbf{} & \multicolumn{1}{c}{\textbf{EER(\%)}} & \multicolumn{1}{c}{\textbf{EER(\%)}} & \multicolumn{1}{c}{\textbf{MinDCF}} \\
    \midrule
    Baseline & 0.72 & 1.21 & 0.0568 \\
    \midrule
     Domain-balanced FT & 0.74 & 1.14 & 0.0523 \\
     LM-FT (m=0.3, s=3) & 0.66 & \textbf{1.08} & \textbf{0.0502} \\
     LM-FT (m=0.5, s=5) & \textbf{0.60} & 1.25 & 0.0573  \\
    \bottomrule
  \end{tabular}
\end{table}

The proposed ECAPA CNN-TDNN architecture outperforms the baseline ECAPA-TDNN model significantly. After fine-tuning and applying quality metrics, the EER and MinDCF on the SdSVC-21 test set show a relative improvement of 11.4\% and 9.4\%, respectively. Similar gains are observed on SdSVC-21 Test for the proposed fwSE-ResNet34 with relative improvements of 7.7\% in EER and 9.2\% in MinDCF.

The proposed LM-FT strategy for the SdSVC-21 target domain is beneficial and produces an average relative improvement of EER and MinDCF of 10.6\% and 11.1\% respectively on the SdSVC-21 test set across the different systems. The SdSVC-21 specific quality-aware calibration improves results on the SdSVC-21 test set further with an average relative improvement of 5\% in both the EER and MinDCF value.

\begin{table}[h]
    \caption{Fusion system performance on SdSVC-21 Test.}
    \label{tab:fusion_results}
  \centering
  \begin{tabular}{lcc}
    \toprule
        \multicolumn{1}{c}{\textbf{Systems}} &
        \multicolumn{1}{c}{\textbf{EER(\%)}} &
        \multicolumn{1}{c}{\textbf{MinDCF}} \\
    \midrule
    Fusion & 1.06 & 0.0473 \\ 
    Fusion + LM-FT & 0.92 & 0.0416 \\ 
    Fusion + LM-FT + QMs & \textbf{0.86} & \textbf{0.0386} \\ 
    \bottomrule
  \end{tabular}
\end{table}

The results of our final fusion submission for SdSVC-21 are shown in Table~\ref{tab:fusion_results}. We also provide the impact of LM-FT and quality-aware calibration. The four system ensemble consists of the baseline ECAPA-TDNN, the proposed ECAPA CNN-TDNN along with a second variant and the proposed fwSE-ResNet34 with positional encodings. See our SdSVC-21 technical report for more details. The weighted score fusion relatively improves upon the strong ECAPA CNN-TDNN single system on the SdSVC-21 test set by 14.9\% and 18.0\% in EER and MinDCF, respectively.

\section{Conclusions}

In this paper we proposed a 2D convolutional stem for the ECAPA-TDNN speaker verification model, incorporating frequency translational invariance in the initial layers of the network. We also introduced frequency-wise Squeeze-Excitation blocks and positional encodings for ResNet architectures, allowing the model to fully exploit positional information on the frequency axis. The proposed modifications significantly improve upon the strong baseline systems. 
Our final SdSVC-21 fusion submission containing the proposed architectures, tweaked by a large margin fine-tuning strategy and a quality-aware calibration step, produces a MinDCF score of 0.0386 and EER of 0.86\% on the SdSVC-21 test set. These promising results encourage us to further explore hybrid architectures in the future.

\bibliographystyle{IEEEtran}

\bibliography{mybib}


\end{document}